\documentclass[12pt]{article}

\usepackage[LY1]{fontenc}               % specify text font encoding
\usepackage{times}                      % switch text fonts (redundant)

\usepackage{graphicx}

\author{{\bf Ray C. Fair}\thanks{Cowles Foundation, Department of Economics,
Yale University, 
New Haven, CT 06520-8281. 
Phone: 203-980-0646; e-mail: ray.fair@yale.edu;
website: {\it fairmodel.econ.yale.edu}.  
I am indebted to 
Rochelle Edge, Refet G\"{u}rkaynak,
Marcin Kolasa, Michal Rubaszek, Pawel Skrzypczy\'{n}ski,
Maik Wolters, Michael Cai, Marco Del Negro, Marc Giannoni, 
Abhi Gupta, Pearl Li, and Erica Moszkowski 
for supplying me with their forecast data. I am also 
indebted to David Childers for helpful comments. 
}}
\date{{\bf August 2018}}
\title{{\bf Information Content of DSGE Forecasts
}}

\begin{document}

\setlength{\unitlength}{.1in}

\maketitle
\thispagestyle{empty}

\begin{abstract}
This paper examines the question whether information is contained in 
forecasts from DSGE models beyond that contained in lagged values, which
are extensively used in the models.
Four sets of forecasts are examined.  The results 
are encouraging for DSGE forecasts of real GDP.  The results suggest that 
there is information in the DSGE forecasts not contained in forecasts based only
on lagged values and that there is no information in the lagged-value 
forecasts not contained in the DSGE forecasts. The opposite is true 
for forecasts of the GDP deflator. 
\end{abstract}

\renewcommand{\baselinestretch}{1.6}\small\normalsize

\section{Introduction}
This paper examines the question whether information is contained in 
forecasts from DSGE models beyond that contained in lagged values.
Lagged variables enter DSGE models through assumptions like habit formation,
adjustment costs, variable capacity utilization, 
pricing behavior, and interest rate rules.  Theoretical
restrictions are imposed on these variables, and the question is whether 
predictive information is added by the restrictions?  

Consider an $s$-period-ahead forecast of real GDP.\ Let $Y^a_t$ denote the 
$s$-period ahead forecast of log GDP for period $t$ from model $a$, 
and let $Y^b_t$ denote the 
same from model $b$.  The forecasts are assumed to be made at the end of 
period $t-s$.  The comparison method used in this paper is discussed 
in Fair and Shiller (FS) (1990).  For the $s$-period-ahead forecasts for 
periods $1$ 
through $T$, the following regression is run:
\begin{equation}
Y_t-Y_{t-s} = \alpha +\beta (Y^a_t-Y_{t-s}) + \gamma (Y^b_t-Y_{t-s}) 
+ u_t, ~~~ t = 1,...,T.
\end{equation}
If neither model contains information useful for $s$-period-ahead 
forecasting of $Y_t$, then the estimates of $\beta $ and $\gamma $ 
should both be zero. In this case the estimate
of the constant term $\alpha $ would be the average
$s$-period-change in $Y$. If both models contain
independent information for $s$-period-ahead
forecasting, then $\beta $ and $\gamma $ should both be
nonzero. If both models contain information,
but the information in, say, model $b$ is 
completely contained in model $a$ and model
$a$ contains further relevant information as
well, then $\beta $ but not $\alpha $ should be nonzero. (If
both models contain the same information,
then the forecasts are perfectly correlated,
and $\beta $ and $\alpha $ are not separately identified.)
It may be that both coefficient estimates are significant, but one is 
negative.  This means that the information contained in the forecast of the 
model with the negative coefficient estimate 
contributes negatively to the overall forecast conditional on the information
in the other model's forecast. 

One model's forecasts 
may have a higher root mean squared error (RMSE) than
another's, but still contain useful independent information.  
Estimating equation (1) allows one to test for this, which the simple 
comparison of RMSEs cannot.   

Further discussion of this method is in Fair and Shiller (1990).
The error term $u_t$ is likely to be heteroskedastic and be a $s-1$ moving 
average process.  This can be corrected for when estimating 
the standard
errors of the coefficient estimates. The 
procedure discussed in Hansen
(1982), Cumby, and  Obstfeld (1983), and 
White and Domowitz (1984) can be used to estimate
the asymptotic covariance matrix of
the coefficient estimates.
When $s$ equals 1 the covariance
matrix is simply White's (1980) correction for heteroskedasticity.   

In the next section a set of comparison rules is suggested.  The forecasts
are discussed in Section 3, and comparison results are 
presented in Section 4.  

\section{Suggested Comparison Rules}
\begin{enumerate}
\item Use a common forecast period.  Some periods are obviously more difficult
to forecast than others, and so a common period is essential.
\item With the exception discussed in point 3, no future information should
be used in making the forecasts.  Rolling estimation can be used up to the 
first period forecast, so no future information is in the coefficient
estimates.  If there are exogenous variables, no future information should be 
used to forecast these variables.  Possibilities are mechanical rules or 
autoregressive equations. In principle future 
information should not be used in calibrating parameters, although this
may be hard to do.  Future information may also have been used in the 
specification the model, since the latest specification is likely to be 
used.  This then means that the forecasts are not true 
ex ante forecasts.  The comparison exercise is conditional on the theoretical
specification of the model and possibly on some calibrated parameters.  
\item Use the latest revised data for the comparisons.  The latest revised
data may also be used for the estimation, which is where future information 
comes in.  The latest revised data are the best estimates of the economy, which
argues for their use.  Also, even 
if real time data are used in the estimation, 
it is not clear what data should be used for the future comparisons.  Using,
say, the first or second estimate of the future data seems worse than using 
the latest data, since one is after the best estimate of the economy.  
Fortunately, as discussed below, using real time versus latest revised data
generally makes only a small difference in the results.  
\item The forecasts should be made by the proprietors of the models.  
Models are complicated, and proprietors know them best.  Allowing an 
outsider to generate the forecasts increases the chances of errors and of 
misrepresenting the model.  
One, of course, has to trust that the proprietors are not cheating, but
programs can be made available to others to duplicate the results.  
\end{enumerate}
  
\section{The Forecasts}
\subsubsection*{DSGE Forecasts}
Four sets of forecasts from DSGE models were used:
Wolters (2013), 
Kolasa, Rubaszek, and Skrzypczy\'{n}ski (KRS) (2012), 
Edge and G\"{u}rkaynak (EG) (2010), and Cai, Del Negro, Giannoni, Abhi, Li,
and Moszkowski (NYFRB) (2018).   
The forecast
periods differ, but the forecasts have all been generated using no 
future information
except for the specification of the model, possibly some calibrated 
parameters, and possibly the use of revised data. Also, for the NYFRB forecasts
the Blue Chip
expectations of the future federal funds rate and the ten-year inflation 
rate are taken as ``data'' during the zero lower bound period.   

Forecasts of real GDP and the GDP deflator have been used from the 
four studies.  In addition forecasts of consumption and investment have 
been used for NYFRB.  
The earliest forecast period is Wolters, 1984:1--2002:4.  
Wolters compares four models, and I have taken the Smet-Wouters model for 
the present analysis.  Wolters uses both real-time data and revised data for
the estimation, and in the spirit of the suggested rules in Section 2, 
I have taken the version using revised data.  Wolters
reports (p. 87) that the relative performance of the models is not sensitive 
to which data are used.  I have also used the forecasts with jump off date 
$-$1.  There are 5 
missing forecasts in the data set, 
so the total number of observations is 63. This forecast period
does not include the housing boom of the early 2000's nor the recession that 
followed.  Forecasts are available for up to 9 quarters ahead.  

For KRS the forecast period is 1994:1--2008:4.  This includes the 
housing boom period, but only the first few quarters of the recession.   
There are 56 observations. Forecasts are available for up to 5 quarters 
ahead.  KRS use real-time data for the estimation of the model, but revised
data for the forecast evaluations---the last vintage data in their sample.
They report (p. 1313) that results using other ``actuals'' are broadly the 
same.  The model is essentially the Smets-Wouters model.  

For EG the forecast period is 1992:1--2010:1, so it does include the
recession.  EG analyze two forecast periods, and this is the longer of the 
two.  They use real-time data for the estimation. For the longer period 
used here the vintage dates are Blue Chip dates.   
Forecasts up to 8 quarters ahead are available, but the 
data are such that there is one fewer observation per quarter ahead.  
There are 73 observations for the 1-quarter-ahead forecast, 72 for the 
2-quarter-ahead, through 66 for the 8-quarter-ahead. 

For NYFRB there are 97 16-quarter-ahead forecasts, with start dates 
1992:1 through 2006:1.  Real-time data are used for the estimation, 
and the model is reestimated once a year. The forecasts are from 
model SWFF, which is the Smets and Wouters (2007) model augmented with
financial frictions.   

The data I have used for the comparisons are revised data as of 
January 26, 2018, which have observations through 2017:4.  
This means for the NYFRB forecasts that errors are available for all 97
8-quarter-ahead forecasts and earler.  For the 9-quarter-ahead 
forecasts 96 errors are available, and so on through the 16-quarter-ahead 
forecasts, where 89 errors are available.  
. 
The forecasts are available from the model builders 
as quarterly percentage changes.
I have converted these forecasts to level forecasts 
using for each variable the actual value (from the revised data) 
on the level of the
variable for the quarter before the first quarter forecast.  Computing 
level forecasts allows one to compare
$s$-period-ahead forecasts for $s$ greater than one.
  
So to summarize, I have taken the exact percentage change forecasts from
the model builders---one set based on estimates using 
revised data (Wolters) and
the other three on real-time data---converted these to levels using the 
latest revised
data for initial starting points, and used the latest revised data for the 
forecast-period comparisons.  

\subsubsection*{Forecasts using only Lagged Variables}
The model used to generate the forecasts of the GDP deflator will be 
denoted PAR4.  This model is a single linear equation, where the left 
hand side variable is the log of the GDP deflator and the right hand side 
variables are the constant term and the first four lagged values of the 
log of the GDP deflator---a fourth-order autoregressive equation.  
172 sets of forecasts were generated.  The beginning quarter for all the
estimations was 1954:1.  The data ended in 2017:4.  
For the first set the end estimation quarter 
was 1974:4, and the forecast period was 1975:1 through 1978:4.  For the 
second set the respective quarters were 1975:1 and 1975:2 through 1979:1,
and so on.  This gave 172 1-quarter-ahead forecasts through 157 
16-quarter-ahead forecasts. These sets were matched to the relevant 
DSGE sets for the comparisons; not all sets were used.  

The model for real GDP uses lagged values of the components of real GDP and
the GDP identity.  It will be denoted YSAR4.  Ten components of real GDP were
chosen, where real GDP is the sum of the ten.  An AR4 equation is 
specified for each component, where the log of a component is taken to 
be a linear function of the constant term and the first four lagged 
values of the log of 
the \renewcommand{\baselinestretch}{1.}\small\normalsize 
component.\footnote{The ten components 
are (all using real NIPA data) consumption of services, consumption of 
nondurables, consumption of durables, residential investment, fixed 
nonresidential investment, inventory investment, exports, imports (with
a minus sign), government
spending (sum of federal and state and local), and a statistical discrepancy
term due to the use of chain-linked data.  For inventory investment and 
the statistical discrepancy, logs were not taken for the component's 
AR4 equation.}
\renewcommand{\baselinestretch}{1.6}\small\normalsize
YSAR4 thus consists of 11 equations, the 10 component equations and the 
GDP identity.  The same estimation procedure was followed for YSAR4 as 
that for PAR4. There are 
no exogenous variables in the model, and so no 
future information is used for the forecasts except for the use of the 
latest revised data. 

As noted in Section 1, for the NYFRB forecasts comparisons are also made for 
consumption and investment.  This requires a little more explanation. 
There are three consumption components in YSAR4: service, nondurable, and 
durable.  Total real consumption is the sum of these three. The model for
consumption, denoted CSAR4, is taken to be the three AR4 equations 
for the components plus the equation summing the three.  The total 
consumption forecasts are then compared to the NYFRB consumption forecasts 
described above.  Remember that the NYFRB level consumption forecasts are 
generated using the actual value of total real consumption in the 
quarter before the forecast begins plus the real consumption growth rates
from the DSGE model. 
In most DSGE models, including NYFRB, real consumption is mismeasured. 
It is taken to be nominal consumption divided by the GDP deflator.  The 
growth rates are thus growth rates of a mismeasured variable, whereas 
the actual initial value of the level of real consumption that is used 
to generate the level forecasts is the correct value. Also, the DSGE level 
forecasts that are generated are compared to the correct actual values of the 
levels (not the mismeasured levels). 
In other words, the assumption here is that the NYFRB growth 
rates of real total consumption are the growth rates of the correct value 
even though they are of the mismeasured value.  

Regarding investment, there are two fixed investment components in YSAR4: 
nonresidential and residential.  Total real fixed investment is the sum
of the two.  The model for
consumption, denoted ISAR4, is taken to be the two AR4 equations 
for the components plus the equation summing the two.  
The total fixed investment forecasts are then compared to the NYFRB 
investment forecasts 
descrbed above.  Similar issues pertain to investment as pertained to
consumption above.  For NYFRB investment is mismeasured as nominal 
investment divided by the GDP deflator, but this is essentially ignored here.

Finally, for NYFRB one can examine the difference between GDP and 
consumption plus investment.  This difference is inventory investment plus
exports minus imports plus government spending plus the statistical 
discrepancy term.  Results are also presented for this difference, denoted 
OTHER, below.  A forecast of OTHER is simply the forecast of real GDP minus
the forecasts of consumption and investment.  

In the estimation of equation (1) in the next section, the 
procedure discussed in Section 1 was use for the estimation of the 
standard errors of the coefficient estimates except for Wolters.  For 
Wolters there are 5 missing observations, and no adjustments were made to 
the OLS estimates of the standard errors.

\section{The Results}
Estimates of equation (1) for the four sets of forecasts 
are presented in Table 1 for real GDP.\  
The quarters ahead analyzed are 2, 4, and 9 for Wolters, 2 and 5 for KRS,
2, 4 and 8 for EG, and 2, 4, 8, and 12 for NYFRB.  

The results in Table 1 are clear: the DSGE forecasts dominate the 
YSAR4 forecasts.  The estimates of $\beta $ are always significant, and only
two of the estimates of $\gamma $ are.  The results thus say that the 
DSGE forecasts of real GDP contain independent 

\renewcommand{\baselinestretch}{1.6}\small\normalsize
\renewcommand{\baselinestretch}{1.}\small\normalsize

\begin{center}

\begin{tabular}{lcccccc}
\multicolumn{7}{c}{\bf Table 1} \\
\multicolumn{7}{c}{\bf Estimates of Equation (1) for Real GDP}\\
\multicolumn{7}{c}{\bf $Y_t-Y_{t-s}$ is the left-hand-side variable.}\\
  [.25ex] \hline \\  [-1.0ex]
\multicolumn{1}{c}{}&
\multicolumn{1}{c}{cnst}&
\multicolumn{1}{c}{DSGE}&
\multicolumn{1}{c}{YSAR4}\\
\multicolumn{1}{c}{$s$}& 
\multicolumn{1}{c}{$\hat{\alpha }$}&
\multicolumn{1}{c}{$\hat{\beta }$}&
\multicolumn{1}{c}{$\hat{\gamma }$}&
\multicolumn{1}{c}{SE}&
\multicolumn{1}{c}{R$^2$}&
\multicolumn{1}{c}{\# obs.}\\
  [.25ex] \hline \\  [-1.0ex]
\multicolumn{7}{c}{Wolters: 1984:1--2002.4}\\
2&0.009&0.332&0.338&0.0079&0.127&63\\
 & (2.10)&(2.34)&(0.90)\\
[.50ex]
4&0.037&0.328&-0.484&0.0134&0.083&63\\
 & (3.82)&(2.21)&(-1.17)\\
[.50ex]
9&0.144&0.293&-1.605&0.0213&0.212&63\\
 & (6.03)&(2.02)&(-3.64)\\
[.50ex]
\multicolumn{7}{c}{KRS: 1994:1--2008:4}\\
2&-0.001&0.657&0.522&0.0066&0.307&56\\
 & (-0.36)&(2.97)&(1.37)\\
[.50ex]
5&0.004&1.010&-0.213&0.0137&0.414&56\\
 & (0.29)&(3.31)&(-0.40)\\
[.50ex]
\multicolumn{7}{c}{EG: 1992:1--2010:1}\\
2&0.709&0.518& 1.102&0.0097&0.308&71\\
 & (2.97)&(3.01)&(2.01)\\
[.50ex]
4&1.262&0.922&0.226&0.0172&0.275&70\\
 & (3.63)&(3.57)&(0.53)\\
[.50ex]
8&2.035&1.502&-0.436&0.0231&0.500&66\\
 & (3.51)&(3.36)&(-0.68)\\
[.50ex]
\multicolumn{7}{c}{NYFRB: 1992:1--2016:1}\\
2&1.552&1.122& 0.379&0.0066&0.590&97\\
 & (6.76)&(6.75)&(1.47)\\
[.50ex]
4&1.504&1.082&-0.038&0.0134&0.426&97\\
 & (4.52)&(4.52)&(0.08)\\
[.50ex]
8&1.280&0.904&-0.298&0.0252&0.250&97\\
 & (3.02)&(2.95)&(-0.53)\\
[.50ex]
12&1.313&0.935&0.026&0.0351&0.203&93\\
 & (1.98)&(1.93)&(0.05)\\
  [.25ex] \hline \\  [-1.0ex]
\multicolumn{7}{l}{$Y$ is the log of real GDP.}\\
\multicolumn{7}{l}{OLS estimates.}\\
\multicolumn{7}{l}{t-statistics are in parentheses.}\\
\multicolumn{7}{l}{Estimated standard errors are 
corrected for heteroscedasticity}\\
\multicolumn{7}{l}{~~~ and a moving average process (except for Wolters).}\\

\end{tabular}
\end{center}

\renewcommand{\baselinestretch}{1.}\small\normalsize
\renewcommand{\baselinestretch}{1.6}\small\normalsize

\noindent  information from that contained in the
lagged values. 
Also, the YSAR4 forecasts contain no information not contained in the 
DSGE forecasts. 

Although not reported in the table, results were obtained where the YSAR4 
forecasts were replaced with forecasts from a simple fourth-order 
autoregressive 
process for real GDP.  In other words, the components of 
real GDP were not used.  In this case the results for the DSGE forecasts 
were even better.  So the positive results for the DSGE forecasts are not
due to something weird about the YSAR4 model.  

The results for the price forecasts are in Table 2, which has the same 
format as Table 1.  The alternative model is simply PAR4, a fourth order 
autoregressive process for the log of the GDP deflator.  In this case the 
DSGE forecasts do not do well.  For Wolters and NYFRB the PAR4 forecasts 
completely dominate.  None of the estimates of $\beta $ are significant, and 
the estimates of $\gamma $ always are. The PAR4 forecasts contain 
independent information from that contained in the DSGE forecasts, and the 
DSGE forecasts contain no information not contained in the 
PAR4 forecasts. 

For KRS and EG the estimates of $\beta $ are 
negative and significant or close to being significant. The estimates of 
$\gamma $ are significant and close to 1.0.  This suggests that conditional
on the PAR4 forecasts, the DSGE forecasts contain additional information,
where the additional information contributes negatively to the overall 
forecast of the GDP deflator. In this case both the DSGE and PAR4 forecasts
contain independent information.  

\renewcommand{\baselinestretch}{1.6}\small\normalsize
\renewcommand{\baselinestretch}{1.}\small\normalsize

\begin{center}

\begin{tabular}{lcccccc}
\multicolumn{7}{c}{\bf Table 2} \\
\multicolumn{7}{c}{\bf Estimates of Equation (1) for the GDP Deflator}\\
\multicolumn{7}{c}{\bf $P_t-P_{t-s}$ is the left-hand-side variable.}\\
  [.25ex] \hline \\  [-1.0ex]
\multicolumn{1}{c}{}&
\multicolumn{1}{c}{cnst}&
\multicolumn{1}{c}{DSGE}&
\multicolumn{1}{c}{PAR4}\\
\multicolumn{1}{c}{$s$}& 
\multicolumn{1}{c}{$\hat{\alpha }$}&
\multicolumn{1}{c}{$\hat{\beta }$}&
\multicolumn{1}{c}{$\hat{\gamma }$}&
\multicolumn{1}{c}{SE}&
\multicolumn{1}{c}{R$^2$}&
\multicolumn{1}{c}{\# obs.}\\
  [.25ex] \hline \\  [-1.0ex]
\multicolumn{7}{c}{Wolters: 1984:1--2002.4}\\
2&0.001&0.094& 0.730&0.0027&0.627&63\\
 & (0.94)&(0.98)&( 9.47)\\
[.50ex]
4&0.002&0.207& 0.641&0.0048&0.637&63\\
 & (0.52)&(1.61)&( 9.12)\\
[.50ex]
9&0.011&0.252& 0.443&0.0121&0.431&63\\
 & (0.81)&(0.99)&( 5.56)\\
[.50ex]
\multicolumn{7}{c}{KRS: 1994:1--2008:4}\\
2&0.005&-0.522& 1.063&0.0027&0.421&56\\
 & (2.74)&(-2.70)&( 4.88)\\
[.50ex]
5&0.017&-0.620& 0.972&0.0055&0.462&56\\
 & (5.23)&(-2.53)&( 3.57)\\
[.50ex]
\multicolumn{7}{c}{EG: 1992:1--2010:1}\\
2&0.003&-0.320& 1.000&0.0030&0.405&71\\
 & (1.49)&(-3.67)&( 4.85)\\
[.50ex]
4&0.008&-0.365& 0.926&0.0055&0.359&70\\
 & (2.11)&(-1.94)&( 3.55)\\
[.50ex]
8&0.032&-0.514& 0.723&0.0098&0.234&66\\
 & (3.56)&(-1.56)&( 2.25)\\
[.50ex]
\multicolumn{7}{c}{NYFRB: 1992:1--2016:1}\\
2&0.004&-0.007& 0.613&0.0031&0.308&97\\
 & (2.81)&(-0.11)&(5.11)\\
[.50ex]
4&0.008&0.009&0.546&0.0053&0.311&97\\
 & (3.58)&(0.11)&(3.96)\\
[.50ex]
8&0.024&-0.021&0.373&0.0101&0.172&97\\
 & (3.76)&(-0.14)&(2.07)\\
[.50ex]
12&0.047&-0.087&0.240&0.0141&0.087&93\\
 & (2.98)&(-0.39)&(2.47)\\
  [.25ex] \hline \\  [-1.0ex]
\multicolumn{7}{l}{$P$ is the log of the GDP deflator.}\\
\multicolumn{7}{l}{OLS estimates.}\\
\multicolumn{7}{l}{t-statistics are in parentheses.}\\
\multicolumn{7}{l}{Estimated standard errors are 
corrected for heteroscedasticity}\\
\multicolumn{7}{l}{~~~ and a moving average process (except for Wolters).}\\

\end{tabular}
\end{center}

\renewcommand{\baselinestretch}{1.}\small\normalsize
\renewcommand{\baselinestretch}{1.6}\small\normalsize

The results for NYFRB for consumption, investment, and OTHER are in 
Table 3, which has the same format as Tables 1 and 2. The results in Table
3 are less systematic across forecast horizons than are those in 
Tables 1 and 2.  For consumption the DSGE estimates of $\beta $ are 
significant or close to being significant across the four horizons.  The 
estimates of $\gamma $ are significant for horizons of 2 and 4 
quarters.  The results thus say that both forecasts contain independent
information for horizons 2 and 4 quarters, but for horizons 8 and 12 
quarters the CSAR4 forecasts contain no information not in the DSGE 
forecasts and the DSGE forecasts contain additional information. 

For horizons of 2 and 4 quarters for both investment and OTHER the 
estimates of $\beta $ are not significant and the estimates of $\gamma $ 
are.  The DSGE forecasts thus do not contain information not in the 
ISAR4 and OSAR4 forecasts, and the latter contain information not in the 
former.  For horizons of 8 and 12 quarters the R squares are quite low.  
The estimate of $\beta $ is significant for investment and horizon 12 and 
for OTHER for horizon 8, but it is hard to know what to make of this.  
Very little is explained for horizons 8 and 12 by either forecast.   

\renewcommand{\baselinestretch}{1.6}\small\normalsize
\renewcommand{\baselinestretch}{1.}\small\normalsize

\begin{center}

\begin{tabular}{lcccccc}
\multicolumn{7}{c}{\bf Table 3} \\
\multicolumn{7}{c}{\bf Estimates of Equation (1) for Consumption,}\\ 
\multicolumn{7}{c}{\bf Investment, and OTHER}\\
\multicolumn{7}{c}{\bf Forecasts are NYFRB Forecasts} \\
  [.25ex] \hline \\  [-1.0ex]
\multicolumn{1}{c}{}&
\multicolumn{1}{c}{cnst}&
\multicolumn{1}{c}{DSGE}&
\multicolumn{1}{c}{AR4}\\
\multicolumn{1}{c}{$s$}& 
\multicolumn{1}{c}{$\hat{\alpha }$}&
\multicolumn{1}{c}{$\hat{\beta }$}&
\multicolumn{1}{c}{$\hat{\gamma }$}&
\multicolumn{1}{c}{SE}&
\multicolumn{1}{c}{R$^2$}&
\multicolumn{1}{c}{\# obs.}\\
  [.25ex] \hline \\  [-1.0ex]
\multicolumn{7}{c}{Consumption: 1992:1--2016:1}\\
2&-0.009&0.252& 1.579&0.0066&0.366&97\\
 & (-1.64)&(1.71)&(4.20)\\
[.50ex]
4&-0.008&0.676&1.073&0.0124&0.348&97\\
 & (-0.60)&(2.10)&(2.34)\\
[.50ex]
8&0.046&1.481&-0.350&0.0202&0.444&97\\
 & (1.47)&(2.72)&(-0.52)\\
[.50ex]
12&0.088&2.115&-0.889&0.0279&0.434&97\\
 & (1.74)&(3.67)&(-1.13)\\
[.50ex]
\multicolumn{7}{c}{Investment: 1992:1--2016:1}\\
2&0.007&-0.292& 1.091&0.0280&0.394&97\\
 & (1.33)&(-1.77)&(5.02)\\
[.50ex]
4&0.020&-0.325&0.964&0.0581&0.207&97\\
 & (1.60)&(-0.84)&(4.47)\\
[.50ex]
8&0.055&0.065&0.243&0.1122&0.003&97\\
 & (1.72)&(0.17)&(0.42)\\
[.50ex]
12&0.084&0.494&-0.363&0.1394&0.133&97\\
 & (1.23)&(4.46)&(-0.48)\\
[.50ex]
\multicolumn{7}{c}{OTHER: 1992:1--2016:1}\\
2&0.005&0.074& 0.951&0.0230&0.292&97\\
 & (1.42)&(1.63)&(5.30)\\
[.50ex]
4&0.012&-0.028&0.621&0.0340&0.241&97\\
 & (1.51)&(-0.40)&(2.92)\\
[.50ex]
8&0.034&-0.200&-0.020&0.0567&0.086&97\\
 & (1.96)&(-2.12)&(-0.09)\\
[.50ex]
12&0.030&-0.075&-0.265&0.0761&0.030&93\\
 & (1.78)&(-0.49)&(-1.01)\\
  [.25ex] \hline \\  [-1.0ex]
\multicolumn{7}{l}{Left hand side variable is $X_t-X_{t-s}$$P$, where 
$X$ is the }\\
\multicolumn{7}{l}{~~~log of consumption, investment, or OTHER.}\\
\multicolumn{7}{l}{AR4 is CSAR4, ISAR4, or OSAR4---see text.}\\
\multicolumn{7}{l}{OLS estimates; t-statistics are in parentheses.}\\
\multicolumn{7}{l}{Estimated standard errors are 
corrected for heteroscedasticity}\\
\multicolumn{7}{l}{~~~ and a moving average process.}\\

\end{tabular}
\end{center}

\renewcommand{\baselinestretch}{1.}\small\normalsize
\renewcommand{\baselinestretch}{1.6}\small\normalsize

\section{Conclusion}
The results in Table 1 are quite strong for the DSGE forecasts of real GDP.  
The forecasts 
seem to contain all the information in the lagged values, at least as 
reflected in the YSAR4 forecasts, plus more.  This is also true for 
consumption in Table 3 for horizons 8 and 12.  For horizons 2 and 4 there 
is information in the CSAR4 forecasts not in the DSGE forecasts.  
For investment and OTHER in Table 3 the results are weaker, where if one
ignores horizons of 8 and 12 quarters (where very little is explained)
there is no independent information in the DSGE forecasts. 
The DSGE results are poor for the GDP deflator in Table 2, especially for 
Wolters and NYFRB, where the DSGE forecasts contain no independent 
information.  

It should be stressed that the comparisons here are only with respect to 
forecasts from a model with only lagged values as explanatory variables.  
There are clearly other forecasts that could be used.  A key 
difficulty in this area is abiding by the rules in Section 2, in 
particular avoiding the use of future information in generating the 
forecasts.  Cai et al. (2018) contains an extensive comparison 
of the NYFRB forecasts with those of others, although their analysis is 
not structured to examine the independent information question in this 
paper.  

In previous writings I have been critical of the DSGE methodology and 
the use of data---see, 
for example, Fair (2012).  From the perspective of a one who works with 
large scale macroeconometric models, there is a 
lack of care in dealing with the data.  It was mentioned above that 
real consumption and investment are mismeasured.  Also, some of the labor 
force and population variables have not been handled correctly.   
And there is too much aggregation of the data.  The behavior of service, 
nondurable, and durable consumption is quite different in the 
macro economy, and much is likely
to be lost in aggregating the three.  Also, plant and equipment investment 
and housing investment behave much differently, and these should not be 
aggregated.  Ignoring imports is also problematic, since the United States is 
far from being a closed economy and import demand is endogenous.  
On the theory side, the theoretical restrictions are very
tight, especially the imposition of rational expectations and the tight use
of the maximization framework.  

For a critic of DSGE models the results in Table 1 for real GDP may thus be 
surprising.  As noted in the Introduction, lagged values are used 
extensively in DSGE models (sometimes in ad hoc ways!), and one might have 
thought that the use of the lagged values is driving the forecast results. 
The results in Table 1 show that this is not the case.  The use of the 
FS comparison method shows that there 
is information in the DSGE forecasts
for real output not in the lagged values.  The puzzle to a critic 
is why the tight 
theoretical restrictions improve the forecasts.   

The main question about the present results is whether 
future information is being used in forecasting real GDP.
For example, some of the parameters in DSGE models are calibrated and fixed
for all the forecasts.  Is future information used in some of the 
calibrations?  For the NYFRB forecasts, is the use of the Blue Chip
expectations in the zero lower bound period cheating in some way regarding 
the information in the model qua model?  Finally, has information on 
the financial crisis led to specification changes in DSGE models that 
help forecast the crisis period even when the rules in Section 2 are 
followed? 

\newpage

\renewcommand{\baselinestretch}{1.6}\small\normalsize
\renewcommand{\baselinestretch}{1.}

\end{document}